\documentclass [psfig,11pt]{article}
\usepackage{setspace,amsfonts,epsfig,amssymb,mathrsfs}
\usepackage{natbib}
\usepackage[margin=1.0in]{geometry}
\usepackage{hyperref}
\usepackage{xcolor}

\begin{document}

\title{Determinism and General Relativity}
\author{Chris Smeenk and Christian W\"uthrich\thanks{We thank Gordon Belot, Juliusz Doboszewski, JB Manchak, the audience at the PSA2010 in Montreal, the New Directions group at Western University, and two anonymous referees for their valuable feedback and exchanges. We acknowledge partial support from the John Templeton Foundation (grants 61048 and 61387).}}
\date{16 September 2020}
\maketitle

\begin{abstract}
\noindent
We investigate the fate of determinism in general relativity (GR), comparing the philosopher's account with the physicist's well-posed initial value formulations.  The fate of determinism is interwoven with the question of what it is for a spacetime to be `physically reasonable'. A central concern is the status of global hyperbolicity, a putatively necessary condition for determinism in GR. While global hyperbolicity may fail to be true of all physically reasonable models, we analyze whether global hyperbolicity should be (i) imposed by {\em fiat}; (ii) established from weaker assumptions, as in cosmic censorship theorems; or (iii) justified by beyond-GR physics.
\end{abstract}

\section{Introduction}
\label{sec:intro}

Two foundational questions one might ask about any physical theory bring out particularly subtle and interesting features of general relativity (GR). First, is GR a deterministic theory? Second, do all mathematical models of the theory represent physically possible spacetimes? There is a tight connection in GR between these two questions, i.e., between an assessment of what spacetimes are physically possible or reasonable and of whether determinism holds. It is this connection that we explore in this essay. 

Determinism holds if specifying the state of a system uniquely fixes its dynamical evolution.  Spacetimes with exotic causal structure raise a distinctive set of questions regarding the status of determinism in GR, differing from those raised by the (in)famous hole argument.  Below we will bypass the hole argument by assuming that the existence of a unique solution `up to diffeomorphism invariance' is sufficient for determinism in GR.  We will consider cases where uniqueness fails regardless of the interplay between metaphysics and diffeomorphism invariance.\footnote{None of this is to deny the interest of the questions raised by the hole argument \citet{EarmanNorton1987}; see also \citet{Stachel2014} for a recent review.  For reasons that will become clear below, we are sympathetic to the position, defended recently by \citet{Weatherall2018}, that we should always take mathematical models to be only individuated `up to isomorphism' for the relevant mathematical objects.}  In the cases we are interested in, fully specifying the state of the universe within a spacetime region fails to fix the maximal extension of the region. Determinism fails in the straightforward sense that everything within the region is compatible with multiple extensions beyond it. Even though the state may in some sense locally determine a unique solution, it fails to fix a unique global, maximal solution. 

Our analysis runs parallel to recent discussions of determinism in Newtonian particle mechanics. Philosophers have considered unusual cases illustrating several distinct ways in which determinism can fail in Newtonian mechanics.  For example, if we allow force functions that are not continuously differentiable, Newtonian dynamics admits multiple solutions for the same initial data.\footnote{Several recent discussions have focused on a simple case:  a particle moving along Norton's dome \citep{Norton03}, a constraint surface that does not have a well-defined second derivative at its apex.  A similar pathology arises for the following force function:  $f(r) = m \sqrt{\ell -r}$ for $r \leq \ell$, and $=0$ for $r > \ell$. Because $f(r)$ is not differentiable at $r = \ell$, Newtonian dynamics fails to admit a well-posed IVF.  See \citet{Fletcher2012} for a thorough review, which notes in passing that the provenance of these problems goes back to Poisson in the early $19^{th}$ century.}  But are these `physically reasonable' forces, falling within the domain of applicability of Newtonian theory? Similarly, many solutions of Einstein's field equation (EFE) have unusual mathematical features that seem unreasonable in some sense, and apparently do not describe the actual universe or a region of it.  How do we decide to treat some solutions as beyond the pale?  As with \citet{Malament2008}'s take on whether Newtonian mechanics is deterministic, our interest is in clarifying the basis for such judgments rather than rendering a verdict on determinism.\footnote {Throughout this paper, we draw on the seminal survey of related topics in \citet{ear95}.}  

Making the world safe for determinism in GR requires substantive claims regarding the global structure of space and time along with constraints on the permissible types of matter.  The conditions needed to reject some solutions as `unphysical' are quite different in character than the local dynamical laws.  Although the content of these conditions is clear, their status is not. We will discuss three different approaches below that reflect different reasons for placing tighter constraints on physical possibility than those implied by EFE alone.  The first approach offers a modern transcendental argument:  although GR dispenses with much `background structure', there are still some global conditions on the spacetime manifold that are prerequisites for sensible physics. In this spirit, conditions closely tied to determinism may turn out to be necessary in order for GR to get any traction at all and then be parlayed into an argument to the conclusion that GR is deterministic. The second approach shifts the goal posts:  rather than attempting to prove determinism in an entirely general sense, cosmic censorship theorems aim to show that a suitably restricted subset of solutions satisfies a stability property.  Since stability under small perturbations is arguably a property of `physically reasonable' spacetimes, this approach offers a second avenue from physical reasonableness to determinism. Finally, a third approach emphasizes that our assessment of `physical possibility' must take into account the limits of applicability of classical GR and the implications of other theories.  We briefly discuss the possibility that a successor theory may rule out the failures of determinism allowed in classical GR, and the related possibility that global conditions may serve as preconditions for the possibility of formulating quantum field theory (QFT). 

The present analysis bears on an important debate in the foundations of physics. One side in this debate considers determinism to be a basic {\em principle} crucial to physicists. Advocates of this stance often take it to be part of clarifying the content of a theory to see what is required to satisfy determinism. The other side insists that there can't be such an {\em a priori} principle and typically treats the various ways of insuring determinism as a possibly illegitimate exercise. Our purpose here is not to argue for a position in this debate, but to clarify the distinctive issues that arise in GR. We agree with \citet{earm04} that those who expect a simple answer from modern physics to the question of whether the world is deterministic or not will be frustrated, as we have also argued elsewhere \citep{earsmewut,smewut,wut10}. 
The foundations of modern physics are simply too rich for such a simple-minded expectation to be borne out. 

We will start in \S\ref{sec:determinism} by fixing the terminology and by discussing the implications of moving from the parabolic partial differential equations of pre-relativistic physics to the hyperbolic ones in relativity. In \S\ref{sec:gr}, we develop the conception of determinism in the context of GR. In \S\ref{sec:global}, we consider whether global hyperbolicity is necessary or sufficient as a condition for determinism in GR. In \S\ref{sec:gh}, \S\ref{sec:censor}, and \S\ref{sec:beyondgr}, respectively, we will discuss three approaches to conceiving of the status of global hyperbolicity as a condition (i) to be stipulated a priori, (ii) to be proved from weaker assumptions within GR, and (iii) to be confirmed --- or infirmed --- from physics beyond GR. Brief conclusions follow in \S\ref{sec:conc}. 

\section{Hyperbolic PDEs and Relativity}
\label{sec:determinism}

Let us start from the idea of determinism as requiring that suitably specifying the physical state of a system in some region of spacetime $S$ uniquely fixes, through the dynamics, the state throughout a larger region $S'$ ($S \subset S'$) --- in its strongest form, throughout the entire spacetime.   Physical theories typically describe dynamical evolution using partial differential equations (PDEs).  In this context, the status of determinism translates into the question of whether the system of PDEs specifying dynamical evolution, supplemented with some further input (information regarding the state of the system in $S$), yield a unique solution throughout $S'$. What form the input must take to lead to a uniquely solvable problem, and the qualitative features of the solutions, differs among the three classes of PDEs --- hyperbolic, parabolic, and elliptic. Over the course of the twentieth century physics has shifted to using hyperbolic PDEs to represent fundamental interactions, with profound consequences for the status of determinism, as the following considerations illustrate.\footnote{Obviously parabolic and elliptic PDEs are still used in a wide variety of modeling contexts.  Here we follow \citet{Geroch1996}, who remarks that ``A case could be made that, at least on a fundamental level, all the `partial differential equations of physics' are hyperbolic --- that, e.g., elliptic and parabolic systems arise in all cases as mere approximations of hyperbolic systems.''}

\begin{figure}[t]
\centering
\epsfig{figure=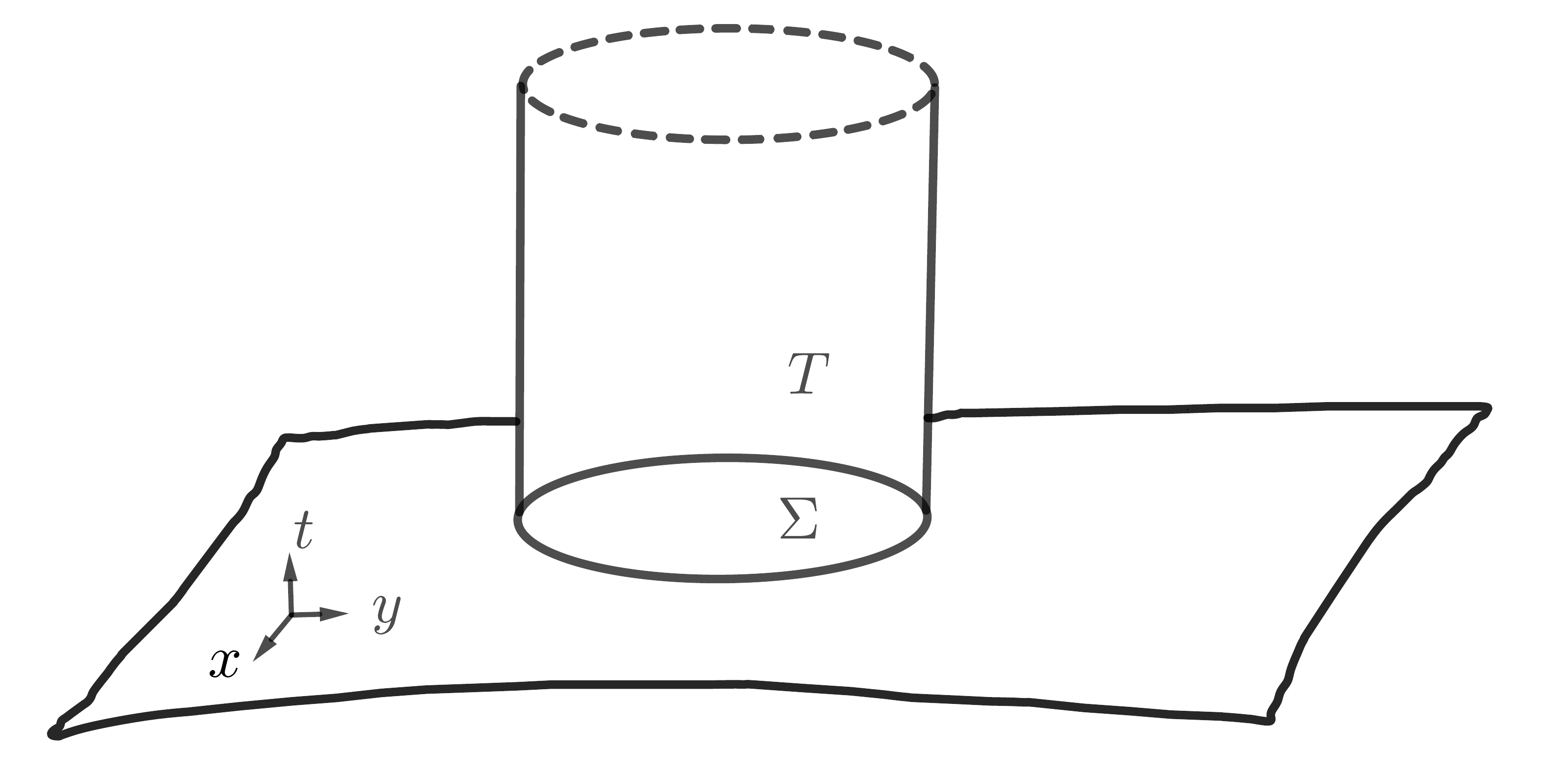,width=0.8\linewidth}
\caption{\label{fig:initialboundary} In this example of an initial-boundary value problem, the region of input data consists of finite region at a given time $\Sigma$ and its cylindrical boundary over time, $T$.}
\end{figure}

One challenge to determinism in Newtonian physics stems from the fact that the relevant equations do not place an upper limit on signal propagation speed.\footnote{Cf.\ \citet{ear86} and \citet{earm04}.} (For Galilean invariant laws, one can always choose an inertial reference frame with respect to which a given particle has an arbitrarily high velocity.)  This raises the possibility of `space invaders':  particles that `fly in from infinity' and disrupt an otherwise well-behaved system.  \citet{Xia1992}  showed that this possibility can be realized in a system of $5$ point particles interacting through Newtonian gravity.  The particles can be arranged such that their trajectories all approach a given time slice asymptotically:  the particles either  show up `out of nowhere', or, considering the time reversed trajectories, `fly off to infinity' within a finite time.  In light of such behavior, the region of spacetime $S$ appropriate for setting up a uniquely solvable problem has to include the boundaries of the system over time -- specifying, for example, that the number of particles remains fixed.  Parabolic PDEs, such as the heat equation, similarly require specifying boundary values to find a unique solution.   For so-called initial-boundary value problems, the region $S$ where input data need to be supplied consists of a region $\Sigma$ at a given time along with its boundary over time (see figure \ref{fig:initialboundary} for an example).

Hyperbolic PDEs have strikingly different qualitative behavior that is a boon to determinism.  The familiar wave equation, $\partial_x^2 u = \partial_t^2 u$, exhibits features that hold for this general class of PDEs.  Solutions are `localized' in that waves produced by a disturbance propagate at finite speed, along so-called characteristic curves, in contrast to the infinite-speed propagation of effects allowed by both parabolic and elliptic PDEs.  The appropriate surface for specifying the input data is a spatially extended region $S$ at a given time.  As a result of the finite propagation speed, there are non-trivial domains of dependence --- roughly speaking, the region of spacetime that can be connected to $S$ by signals going at or below the wave propagation speed.  For any point falling within the domain of dependence, all waves propagating to or from the point must `register' on $S$.  The state of the system throughout this larger region is fixed by the wave equation along with the initial data on $S$. There is no need to supplement the data on $S$ with conditions regarding how the solution behaves `on the edge' of $S$ or at other times.  Maxwell's equations share these features of the wave equation (albeit with additional complications due to gauge freedom).  Taking the causal structure of these equations to be universal rather than characteristic of electromagnetism leads to special relativity.  Minkowski spacetime provides a natural setting for formulating hyperbolic PDEs, as the causal structure of the spacetime reflects the existence of a maximum propagation speed.         

We need to introduce some technical machinery to make this more precise.  A \emph{spacetime} is an ordered pair $\langle \mathcal{M}, g_{ab}\rangle$, where $\mathcal{M}$ is a connected four-dimensional differentiable manifold without boundary, and $g_{ab}$ is a Lorentz-signature metric defined everywhere on $\mathcal{M}$ that satisfies EFE -- there is a stress-energy tensor $T_{ab}$ and value of the cosmological constant $\Lambda$ such that $R_{ab} - \frac{1}{2}Rg_{ab} - \Lambda g_{ab} = 8\pi T_{ab}$.  This last clause only has some bite if there are further restrictions on $T_{ab}$.  Some results regarding causal structure follow without placing any constraints on $T_{ab}$.  Most of the discussion below will focus on either exact solutions, with a specific choice of $T_{ab}$ representing particular type(s) of matter fields, or `generic' solutions for which $T_{ab}$ satisfies constraints such as the energy conditions \citep[see also][159-161]{Malament2012}.  

The metric $g_{ab}$ fixes a light cone structure in the tangent space $\mathcal{M}_p$ at each point $p \in \mathcal{M}$.  A tangent vector $\xi^a \in \mathcal{M}_p$ is classified as timelike, null, or spacelike, according to whether $g_{ab}\xi^a\xi^b >0, =0,$ or $<0$ (respectively). Geometrically, the null vectors `form the light cone' in the tangent space with the timelike vectors lying inside and the spacelike vectors lying outside the cone.  The classification of tangent vectors extends to curves, and for a time orientable spacetime, which admits a globally consistent designation of the `future' lobe of the null cone, one can further define future-directed timelike (causal) curves as those whose tangent vectors fall within (on or within) the future lobe at each point. 

The light cone structure and classification of curves leads directly to more general characterizations of the causal structure, including domains of dependence.  The \emph{chronological past} $I^-(p)$ is defined as the set of all points in $\mathcal{M}$ that can be reached from $p$ by past-directed timelike curves (analogously for the chronological future $I^+(p)$). For an \emph{achronal surface} $V$, there are no points $p,q \in V$ such that $p \in I^-(q)$.  The \emph{edge} of an achronal surface $V$ is the set of points $p$ such that every open neighborhood $O$ of $p$ includes points in $I^{+}(p)$ and $I^{-}(p)$ that can be connected by a timelike curve that does not cross $V$.  A \emph{global time slice} of a spacetime $\langle \mathcal{M}, g_{ab}\rangle$ is a closed, achronal hypersurface $\Sigma\subset \mathcal{M}$ with no edges. The \emph{future domain of dependence} $D^+(\Sigma)$ for a global time slice $\Sigma$ includes the points in $\mathcal{M}$ such that every past inextendible causal curve through $p$ intersects $\Sigma$ (analogously for the past domain of dependence $D^-(\Sigma)$).\footnote{A past inextendible causal curve $\gamma(s): I \rightarrow \mathcal{M}$ does not have a past endpoint, namely a point $p \in \mathcal{M}$ such that for every neighborhood $O$ of $p$, there is a value $s_0 \in I$ such that $\gamma(s) \in O$ for all values $s<s_0$.}  A spacetime $\langle \mathcal{M}, g_{ab}\rangle$ is \emph{globally hyperbolic} just in case there is a slice $\Sigma$ such that the domain of dependence $D(\Sigma):=D^+(\Sigma) \cup D^-(\Sigma)$ is the entire spacetime, i.e., $D(\Sigma) = \mathcal{M}$.  

With these definitions in place, the contrast between Newtonian physics and special relativity can be stated simply:  input data on a time slice $\Sigma$ in Newtonian spacetime gets us nowhere, in the sense that $D^+(\Sigma) = D^-(\Sigma) = \Sigma$, whereas Minkowski spacetime allows for the possibility of non-trivial domains of dependence.  Whether this possibility is realized depends on whether the matter fields being studied respect the causal structure.  In particular, matter fields that satisfy the \emph{dominant energy condition} do not propagate outside the light cone.\footnote{A matter field with stress-energy tensor $T_{ab}$ satisfies the dominant energy condition if $ T_{ab} \xi^a \xi^b \geq 0$ and $T^a_b \xi^b $ is timelike or null, given an arbitrary timelike vector field $\xi^a$ at any point in $\mathcal{M}$ \citep[see, e.g.][144f]{Malament2012}.}  Proofs that matter fields admit a well-posed initial value formulation (IVF) in Minkowski spacetime depend on what are called `energy estimates'. Consider slices $\Sigma_1$ and $\Sigma_2$, at a later time, and the regions $S \subset \Sigma_1$ and $T = D^+(S) \cap \Sigma_2$.  For matter fields satisfying the dominant energy condition, the field equations imply bounds on the energy associated with initial data on $S$ and $T$.\footnote{The energy over a spacelike region $S$ is obtained by integrating $T_{ab}\xi^a\xi^b$ over $S$, where $\xi^a$ is a timelike vector field normal to $S$.  Often the `energy' used in PDE techniques does not, however, correspond directly to a physically meaningful quantity.}  To establish uniqueness of a solution to a linear wave equation for specified initial data, schematically, suppose for \emph{reductio} that there are two solutions with the same initial data on $S$.  Consider a third solution defined as their difference. The energy of this third solution on $S$ is zero, by hypothesis, but it follows from the energy bounds that this quantity cannot differ from zero on $T$ for an arbitrary later slice $\Sigma_2$.  Hence the two solutions must be identical within $D^+(S)$.  Energy estimates also play an essential role in establishing the other two properties required for a well-posed IVF, namely that solutions exist and vary continuously with the initial data.\footnote{See \S 7.4 of \citet{hawell} for a proof that quasi-linear (linear in the highest derivatives), second-order hyperbolic systems admit a well-posed IVF. \label{IVF} The dominant energy condition is not necessary:  in some cases when $T_{ab}$ fails to satisfy this condition, it still satisfies energy inequalities sufficient to prove that there is a well-posed IVF.}   For linear wave equations, with suitable initial data specified on a time slice $\Sigma$ (aka a Cauchy surface), $D(\Sigma) = \mathcal{M}$.\footnote{The proofs for linear wave equations have been generalized to some non-linear systems, but it is much more challenging to prove global existence theorems for these cases and results are limited.  Recent work has also focused on proving theorems given weaker assumptions about the regularity of the initial data and solution spaces \citep[see][]{Rendall08,rin09}.}  Hence the transition to Minkowski spacetime supports the claim that fundamental physics satisfies determinism. 

The consequences of the further step from special relativity to GR are more challenging to assess.  To represent the gravitational field via spacetime geometry, the light cone structure of Minkowski spacetime has to be demoted from global to local status.  Modern presentations of GR sometimes generalize this claim:  the mathematical structures typically treated as global, background features of spacetime prior to GR should be `stripped away', leaving only a bare differentiable manifold.  In GR the causal structure is itself dynamical rather than fixed, leaving us unable to appeal to various background structures used in the analysis of other PDEs.  A further source of complexity is the non-linearity of EFE.  Generally speaking, for non-linear field equations `good' initial data may lead to singularities, where the derivatives `blow up' and existence and uniqueness results fail within a finite time --- as illustrated by the formation of shock fronts in fluid mechanics \citep{Christodoulou2007}.  This possibility poses a general obstacle to understanding the `global' behavior of solutions even if the dynamics admits a well-posed IVF `locally' (close to the initial data surface, in a precise sense).  In addition, although we noted above that we will set aside issues related to the hole argument, the diffeomorphism invariance of GR leads to complications in providing an IVF.  One complication resembles that faced by theories with gauge freedom, such as Maxwell's equations:  the initial data cannot be freely specified but must satisfy a set of constraint equations, elliptic PDEs.    All existing proofs of the existence and uniqueness of solutions for EFE utilize additional structures or coordinate conditions, rather than employing a framework which maintains manifest diffeomorphism invariance throughout. Techniques called hyperbolic reduction, for example, can be employed to derive a system of hyperbolic PDEs from EFE.  As with Maxwell's equations, these have to be accompanied by elliptic constraint equations, and one has to furthermore ensure that the constraints propagate (that is, that they continue to hold under the `time evolution' equations extracted from EFE).  Yet other formulations reduce the EFE to a system of parabolic equations, or lead to constraint equations that are ordinary differential equations (as in the characterisitic initial value problem).\footnote{For further discussion of the different formulations, see, e.g., \citet{friren,rin09}.}  

These aspects of GR combine to make the question of determinism so intriguing.  There is much more work in mathematical physics needed to elucidate the structure of second-order non-linear PDEs such as EFE. An understanding of the status of determinism in GR comparable to what we have achieved in other physical theories depends on the outcomes of this work.  But even as we await resolution of these issues, we might aim as a modest first step to distinguish between failures of determinism that arise due to what looks like a mistake in mathematics, namely stripping away structures that turn out to be essential to formulating physics, and those that reflect underlying features of the dynamics, such as the formation of `naked singularities'.

\section{Defining Determinism}
\label{sec:gr}

With the terminology and the background in place, let us consider various proposals for defining determinism in the context of GR.  \citet[22]{budsac} define a spacetime to be deterministic just in case $\forall p\in \mathcal{M}$, every past-endless causal curve which intersects $I^+(p)$ also intersects $I^-(p)$.\footnote{A causal curve is \emph{past-endless} just in case there is no event $p\in \mathcal{M}$ such that going in the past direction, for every neighbourhood $U(p)$ of $p$, the curve enters and remains in $U$.} As Budic and Sachs themselves admit, this is a very restrictive definition; they even prove that no asymptotically flat spacetime is deterministic in their sense. Among many other important models, it classifies Minkowski spacetime as non-deterministic. However, they defend their characterization as adequate because it captures the intuitive sense that in deterministic spacetimes, ``each event can, in principle, predict from its own past everything $[$i.e., its chronological future$]$, not merely its own future'' (ibid., 23). 

We do not share this intuition. In fact, we find it curious that anyone would think that the past of a single event would suffice to fix its complete (chronological) future in a dynamical system where effects propagate at finite speed, i.e., in a system described by hyperbolic PDEs! Our position here is supported by results which have become known under the label of `observational indistinguishability', to be discussed in \S\ref{sec:gh}, which establish that the past of an event does not, quite generally and under relatively mild assumptions, fix the rest of the spacetime.  Furthermore, the proposal blurs the crucial distinction between predictability and determinism.\footnote{This is not to deny the interest of questions regarding predictability.  See, e.g., \citet{ger77}.}  On our view, the fate of determinism does not depend on whether the input data needed to specify a uniquely solvable problem are accessible to an embedded observer.

The conception of determinism we are seeking instead follows the standard Laplacian notion, at least in spirit.\footnote{\citet{Butterfield1989} offers a relativistic conception of determinism, further generalized by \citet[11]{Dobo2019}.  Although we heartily endorse \citet{Dobo2019}'s recognition that there are a variety of further conditions that might be imposed in defining determinism, we hold that there is still a core conception underlying these various modifications.} The Laplacian notion encodes a sense of metaphysical determination and is deprived of epistemic undertones. It captures the idea that given the state of the physical system at stake at a time, the laws of nature as encoded in the theory describing the system uniquely determines the state of the system at all other times.  More precisely, for a globally deterministic theory, for appropriate initial data specified within an appropriate spacetime region $S$ there is a unique solution of the relevant equations (including EFE and equations governing any coupled matter fields) through the entire spacetime ($D(S) = \mathcal{M}$). A more rigorous definition would require filling in the two instances of `appropriate' in our characterization.  For vacuum solutions, the initial data for gravitational degrees of freedom typically include the induced three-dimensional metric $h_{ab}$ and the extrinsic curvature $k_{ab}$ on a three-dimensional hypersurface.\footnote{Provided that they satisfy the Gauss-Codacci constraints.  Although the most familiar IVF for EFE uses initial data of this type, there are other possibilities.  (We will mention one of these below; see \citet{Sarbach2012} for discussion and further references.)}  More generally, the form initial data must take in order to be `appropriate' will depend on the theory of matter to be used.\footnote{For typical field equations, the initial data include the field values and their first derivatives (with respect to the `time' coordinate orthogonal to the surfaces $S$). There are further choices to be made regarding what function space the initial data belong to, based on the degree of regularity required (for both metric and matter degrees of freedom). Oddly, proofs of uniqueness require a higher degree of differentiability than the existence results.}  For hyperbolic PDEs in special relativity, initial data can be assigned to spacelike hypersurfaces of constant time, so a natural candidate for the `appropriate' region is the relativistically next best thing to a moment of time --- that is, a global time slice $\Sigma$.  (But all time slices are not created equal.  We further need to rule out surfaces such as an achronal surface $\Sigma$ for which $\exists p: \Sigma \subset I^-(p)$ in Minkowski spacetime; for such a surface, $D(\Sigma)$ is not the entire manifold, but this reflects a poor choice of $\Sigma$.)  

A theorem due to \citet{ChoquetGeroch1969} establishes that appropriate initial data $(\Sigma, h_{ab},k_{ab})$, specified on a global time slice $\Sigma$, determine a unique (up to isometry), maximal globally hyperbolic development, a solution to vacuum EFE.  If matter fields are included, the initial data will consist of two types of information, specifying the state of the spacetime geometry and the matter fields, respectively, and the dynamical equations at play will then include EFE as well as the dynamical equations of the matter degrees of freedom.  It is generally expected that matter fields that have a well-posed IVF in special relativity also have one in this context.\footnote{The proof mentioned in footnote \ref{IVF} carries over to the coupled Einstein-matter field equations for many types of matter fields, granting some plausible assumptions regarding the coupling between the matter fields and gravitational degrees of freedom. But there are numerous open questions, related for example to the treatment of fluid bodies and the Einstein-Boltzmann equation. See \citet{Rendall08} for further discussion.} 
The restriction of the result to the globally hyperbolic development is as essential as it is unsurprising: regions that are not globally hyperbolic fall outside of $D(\Sigma)$ by definition.  The solution determined by the initial data is the maximal globally hyperbolic development of the data, but it is not the maximal development \emph{simpliciter} (a point we will return to below). 

We should pause to note a subtle contrast between the treatment of an IVF for general relativity and for other PDEs.   The background spacetime structure is not given \emph{ab initio}, and we currently lack mathematical tools that would allow for a fully diffeomorphism-invariant treatment.  What we have instead are a variety of methods for reducing the full EFE to a system of constraint and evolution equations.  Choquet-Bruhat discovered that existence and uniqueness results from hyperbolic PDEs could be extended to EFE, which then take the form of a non-linear wave equation, given initial data $(\Sigma, h_{ab},k_{ab})$ satisfying constraint equations. But this data set $(\Sigma, h_{ab},k_{ab})$ has to be regarded as defined on an `abstract' space --- we cannot, without begging the question, assume that $\Sigma$ fits within a solution.  The existence proof establishes that there is an embedding map $\phi: \Sigma \rightarrow \mathcal{M}$ such that $h_{ab},k_{ab}$ appropriately match the spacetime metric $g_{ab}$, where $\langle \mathcal{M},g_{ab} \rangle$ is a solution to EFE.  Turning to uniqueness, we can further ask whether the same initial data set can be embedded in two distinct solutions. The proof of uniqueness establishes, roughly put, that any such pair of solutions must be extensions of one underlying solution \citep[see, e.g.,][]{rin09}.  This apparatus implies a direct response to the hole argument.  What these arguments establish is often called `geometric uniqueness' in the mathematical physics literature, to indicate that an equivalence class of spacetimes has the `same geometry'.  But it is clear from the way that existence and uniqueness results are formulated that we cannot even pose a threat to determinism without assuming some further structure.  These results establish the existence of embedding maps between manifolds (an abstract `initial data' space and solutions to EFE), and in this context the solutions are only individuated up to isomorphism.  There are no grounds to draw a further distinction between `hole diffeomorphs'.       

Applied mathematicians usually impose a third condition to have a well-posed initial value formulation:  the solutions must depend continuously, in a sense to be made precise, on the initial data.\footnote{A map $\gamma:  \mathcal{X} \rightarrow \Gamma$ between initial data $\mathcal{X}$ and solutions $\Gamma$ is continuous if the inverse images under $\gamma$ of all open sets in $\Gamma$ are open sets in $\mathcal{X}$. Intuitively, points that are nearby in $\mathcal{X}$ are mapped into nearby points in $\Gamma$, such that small changes in initial data lead to small changes in the solution.} Philosophers have for the most part not equated determinism with the existence of a well-posed IVF because they reject this third condition. Several alleged failures of determinism depend on exact specification of a set of initial conditions. In Xia's `space invaders' example mentioned above, the motion of the five particles has to be choreographed with exquisite precision such that they all fly off to infinity.  More precisely, the set of initial conditions that lead to the desired behavior is Lebesgue measure zero in the space of initial data \citep{Xia1992}. Initial conditions `arbitrarily close' to those leading to the space invaders trajectory have quite different dynamical behavior, violating the continuity requirement.\footnote{Xia obtains the initial conditions through a Cantor-set like construction:  for each possible interaction among the particles, eliminate the initial conditions that do not lead to the desired trajectory.  Iterating this process leads to a measure zero set of initial conditions, with trajectories confined to the unstable submanifold corresponding to space invaders behaviour.}  

This third condition raises a number of more subtle questions than the requirements of existence and uniqueness.  In particular, the assessment of continuity and stability depends upon the topology assigned to the space of initial data and the solution space.  Yet the grounds for choosing the topology are not always clear.  \citet{fle16} considers stability questions regarding full spacetimes rather than initial data --- that is, in terms of a topology on the space of Lorentzian metrics over a fixed manifold.  He argues that no proposed candidate for a `canonical' topology on this space yields intuitively correct verdicts on convergence of sequences, continuity of families of metrics, and the stability or genericity of global properties.  He advocates instead a version of reflective equilibrium, according to which the choice of topology is guided by contextual features of specific applications.  For the initial value problem, the context requires a characterization of the initial data and a choice regarding what counts as a solution to the PDE.\footnote{The initial data and solutions are typically required to be elements of Sobolev spaces (spaces of functions whose partial derivatives up to some specified order are square integrable).  The relationship between the function spaces for the initial data and solutions are partially dictated by the PDEs themselves, and partially by the choice of desirable properties of solutions.}  
These decisions may limit the choice of topology in specific applications, perhaps even leaving us with only one viable candidate, although the physical significance of the topology and the justification for these choices deserves further scrutiny.   Despite these open foundational questions, the third requirement strikes us as a plausible candidate for a principle that is required for successful applications of differential equations.  We are finite epistemic agents, unable to determine initial conditions, or carry out the numerical analysis of PDEs, with indefinite precision.  A stability requirement is thus essential to using PDEs in the study of natural phenomena.  Most philosophers, by contrast, seem to allow for indefinitely precise individuation of solutions in constructing instances of indeterminism.  On this view, the existence of a well-posed IVF is sufficient for determinism, but it is not necessary.   In any case, we will follow the philosophical tradition below in equating determinism with only the first two conditions of an IVF --- the existence and uniqueness of solutions to the relevant equations --- while also acknowledging the relevance of stability considerations.

\section{Global Hyperbolicity}
\label{sec:global}

The seminal result due to Choquet-Bruhat and Geroch mentioned above depended on identifying a property of spacetimes necessary to complete the proof:  global hyperbolicity.\footnote{Earlier studies led to the formulation of various conditions needed to ensure the existence of an IVF for hyperbolic PDEs; global hyperbolicity is an extension of these earlier conditions to the context of general relativity.} For such spacetimes, this result establishes that EFE admit a well-posed IVF, and in that sense settles the status of determinism.  But not all spacetimes are globally hyperbolic.  This property is in fact the top of a hierarchy of restrictions on the global causal structure of $\langle \mathcal{M}, g_{ab}\rangle$.  There are solutions of EFE in which this property, and properties further down in the hierarchy, fail to hold.  Perhaps the most famous of these is G\"odel spacetime, which does not contain a global time slice $\Sigma$ and has closed timelike curves (CTCs) passing through every point.  Such spacetimes are causally so pathological that the very application of the concept of determinism remains unclear.  But there are cases in which we can assess determinism when global hyperbolicitiy does not hold.  Here we will consider briefly two very different examples, which challenge whether globaly hyperbolicity is necessary or sufficient for determinism in GR.

Global hyperbolicity may not be a necessary condition for determinism. The discussion above restricted attention to the type of initial value problem natural for hyperbolic equations:  initial data specified on a slice $\Sigma$, with no need to specify boundary conditions on a timelike surface.  But EFE allows for quite different set-ups, including initial-boundary value problems with input data specified on a cylindrical surface $\Sigma \cup T$ (consisting of a spacelike `end cap' $\Sigma$ and timelike sides $T$, see figure \ref{fig:initialboundary}).  \citet{FriedrichNagy} formulated the initial-boundary value problem for vacuum EFE, establishing that appropriate initial data on a surface $\Sigma \cup T$ determine the solution within the cylinder.  This new setting leads to significant technical complications. In particular, there is not a clean separation between the dynamical evolution of the initial data and the constraints imposed on $\Sigma$, and further conditions have to be satisfied where $\Sigma$ and $T$ meet.  Establishing whether two sets of initial data generate the same `spacetime geometry' is also more challenging. Yet this set-up is particularly useful for problems in numerical relativity.\footnote{See \citet{ReulaSarbach2010} for an overview.}     

It is possible to formulate a well-posed initial-boundary value problem for EFE with remarkably weak assumptions about causal structure. Anti-de Sitter (AdS) spacetime is the maximally symmetric solution with $\Lambda < 0$, often studied in string theory.  The light cones `broaden out', such that the solution fails to be globally hyperbolic. Given a slice $\Sigma$ and a point $p$ in $I^+(\Sigma)$, there are past directed causal curves from $p$ that do not intersect $\Sigma$.  This suggests, just as in the example of Newtonian space invaders mentioned above, that initial data on a slice $\Sigma$ do not fix future evolution due to the possibility of a disturbance coming `from infinity'.  Despite this bizarre causal structure, there is evidence supporting the conjecture that EFE admit a well-posed initial boundary value problem in AdS for suitable boundary data on a timelike boundary.\footnote{See \citet{Holzegeletal2015} for a statement of the conjecture, discussion of evidence for asymptotical stability of AdS to linearized perturbations with dissipative boundary conditions, and further references.} The conjecture holds that AdS is asymptotically stable for optimally dissipative boundary conditions, with maximal energy flux through the timelike boundary.  (By contrast, AdS appears to be unstable for the other extreme boundary condition --- namely, a reflecting boundary with vanishing flux.)

In terms of the implications for determinism, these types of solutions lead to questions similar to those raised by the space invaders example in Newtonian physics.  In the Newtonian case, do we take the equations to describe a system with a fixed number of interacting particles?  The affirmative answer often assumed in applying the theory implicitly imposes boundary conditions on a timelike surface.  Similarly, treating gravitational degrees of freedom in AdS as part of an initial boundary value problem leads to a well-posed problem only if we impose further conditions on the timelike boundary.  In this case, the appropriate boundary conditions require that the system constantly ``leaks'' energy associated with the gravitational degrees of freedom, preventing the amplification of perturbations reflecting from the boundary.  Obviously this boundary condition does not itself follow from the dynamics, but has to be, in effect, imposed by hand as part of setting up the problem.  The result shows that, at least in this specific case, the threat to deterministic evolution left open by the failure of global hyperbolicity can be countered by a global stipulation with a strikingly different character.  We do not know whether other non-globally hyperbolic spacetimes also admit well-posed initial-boundary value problems.  In any case, although we will not pursue the point in detail here, this case raises similar questions to those we discuss below regarding global hyperbolicity.  

Let us turn now to the question of whether global hyperbolicity is sufficient to guarantee a form of determinism.  To state the problem clearly we need to first introduce the idea of an extension:  a spacetime $\langle\mathcal{M}, g_{ab}\rangle$ can be {\em extended} if it is isometric to a proper subset of another spacetime.\footnote{That is, it can be extended if there exists a spacetime $\langle\mathcal{M}', g'_{ab}\rangle$, $\mathcal{M}\subsetneq \mathcal{M}'$, and an isometric embedding $\phi:\mathcal{M}\rightarrow \mathcal{M}'$ such that $\forall p\in\mathcal{M}', \phi^\ast(g_{ab}(\phi^{-1}(p))) = g'_{ab} (p)$.  We can further refine this definition by specifying the differentiability class of the isometries.} An extension of a spacetime is {\em maximal} if it is {\em inextendible}, i.e.\ it is not isometric to a proper subset of another spacetime.  The result cited above establishes that appropriate initial data fix a solution throughout the domain of dependence (up to isomorphism); but are they sufficient to establish global uniqueness?  Suppose that the domain of dependence has a future boundary, called the \emph{Cauchy horizon}, $H^+(\Sigma)$; what determines evolution in the regions beyond this boundary?

\begin{figure}[t]
\centering
\epsfig{figure=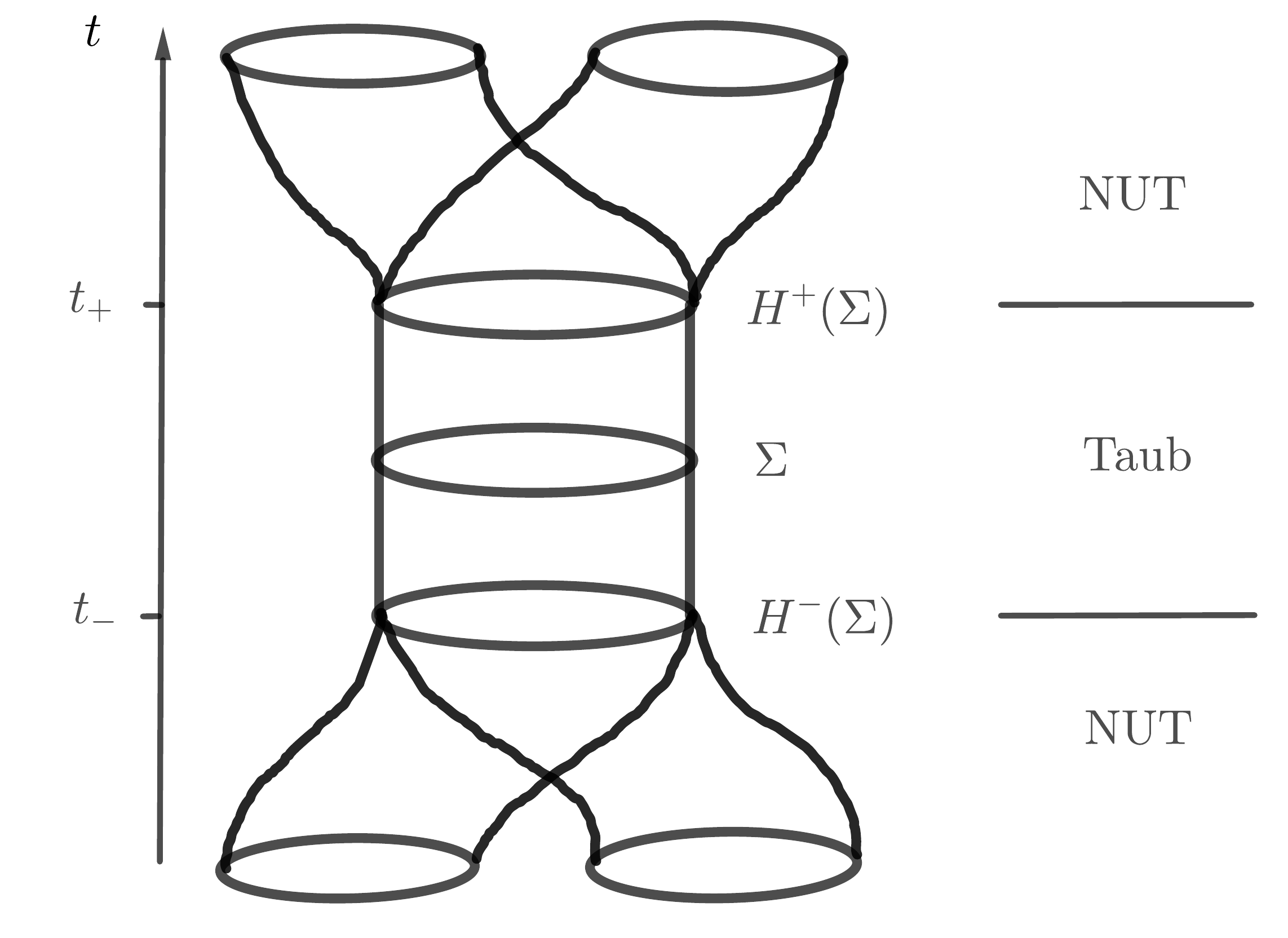,width=0.75\linewidth}
\caption{\label{fig:taubnut} In Taub-NUT spacetime, the Taub region has two extensions (NUT regions) each to the future and the past. Although the figure shows all these extensions at once, we only mean to highlight the distinct possibilities, and not that the spacetime is itself in some sense `branching'.}
\end{figure}

Taub-NUT spacetime, a solution of the vacuum EFE, illustrates a failure of global uniqueness.  One region of this spacetime has global slices $\Sigma$ with a non-trivial domain of dependence (the {\em Taub region} of the spacetime), which is thus globally hyperbolic; however, $D(\Sigma)$ is not maximal and it admits of multiple extensions (the {\em NUT regions}) --- see figure \ref{fig:taubnut}.  Topologically the Taub region is $S^3 \times (t_-,t_+)$, bounded by Cauchy horizons at $t_-$ and $t_+$ ($\in \mathbb{R}$) generated by closed null geodesics.  For any point in the Taub region, there are complete null geodesics that cross the horizon, but there are also incomplete null geodesics that spiral towards the Cauchy horizon.  Despite this singular behavior, the Taub region admits two extensions in both directions -- that is, through $H^+(\Sigma)$ and $H^-(\Sigma)$.  The sense in which the existence of these extensions leads to a failure of uniqueness is surprisingly subtle, however \citep{chrusciel1993nonisometric}:  the two extensions through either horizon are \emph{locally} isometric, but combinations of the different extensions through both horizons can generate non-isometric spacetimes. To illustrate this, let us call the bottom two extensions in figure \ref{fig:taubnut} before $t_-$ `legs' and the top ones after $t_1$ `arms'. The two legs are locally isometric, and so are the two arms. However, the spacetime resulting from adding, say, the left arm and the left leg is not isometric to the one resulting from adding the left arm and the right leg. 

\citet{chrusciel1993nonisometric} generalize this result based on the study of two other families of exact solutions with Cauchy horizons:  there appear to be multiple non-isometric extensions in cases where, as in Taub-NUT, the Cauchy horizon consists of disconnected components. In these cases the failure of uniqueness has a striking global character, since it only arises in considering extensions through disconnected components of the Cauchy horizon.  There is further work to be done to establish that uniqueness also fails due to local extensions through a Cauchy horizon.

If such cases are deemed to be physically possible, then even though EFE locally admits a well-posed IVF this cannot be extended to prove \emph{global} existence and uniqueness.  We will explore below various different stances on whether this simple example can be ruled out in some way.  The extensions may necessarily exploit symmetries of these exact solutions, running afoul of the requirement of stability, or perhaps the unusual causal structure of these models is objectionable. But we should emphasize here that these different responses share one feature, namely that they introduce a definition of physical possibility that goes beyond imposing the local dynamical laws. In fact, this is necessarily so, as the Taub-NUT spacetime and similar examples are solutions of EFE.

We now move to the discussion of the three approaches of saving determinism in GR, focusing on global hyperbolicity as an at least necessary condition for determinism. These three approaches conceive of the status of global hyperbolicity as a condition (i) to be stipulated a priori, (ii) to be proved from weaker assumptions within GR, and (iii) to be confirmed --- or infirmed --- from physics beyond GR.

\section{Global Hyperbolicity by \emph{Fiat}}
\label{sec:gh}

Global hyperbolicity is the primary condition necessary to insure determinism in GR (modulo the concerns in the previous section). The first approach to explicate the status of the global condition of global hyperbolicity essentially {\em stipulates} its validity. It is based on a transcendental argument in that its imposition by {\em fiat} acts as a condition for obtaining a sensible physical theory. This kind of argument is not without precedent: We routinely ask our theories to satisfy certain conditions dictated by the mathematics. For instance, in classical mechanics, the Lipschitz condition is imposed in order to satisfy an antecedent condition of important existence and uniqueness theorems for the PDEs capturing the standard IVF. Thus, its violation leads to forms of indeterminism as exemplified in Norton's dome. 

In spacetime physics, one routinely requires that the manifold be a {\em Hausdorff space}, i.e.\ a topological space such that any two distinct points have disjoint neighborhoods.\footnote{There are other routine requirements with a similar character, for example that spacetime must be ``as large as it can be'' or free of holes.  We do not have space to consider hole-freeness conditions in detail here, but see \citet{Manchak2016epistemic,Dobo2020} for excellent recent discussions.} Typically, demanding that a spacetime manifold be a Hausdorff space is not considered a stipulation with all the advantages of theft over honest toil, but instead a refinement of what is needed mathematically to formulate a sensible theory. Roger \citet[595]{penrose1979} affirms this when he writes that ``I must... return firmly to sanity by repeating to myself three times: `spacetime is a {\em Hausdorff} differentiable manifold; spacetime is a {\em Hausdorff}...' '' The need to refine our definition of a manifold only became apparent with the recognition of bizarre non-Hausdorff cases, e.g., the cases of branching spacetimes considered, but rejected, by Penrose. Although explicitly defended e.g.\ in \citet{belnap1992} and \citet{mccall1994}, John \citet{earman2008} argues that actually branching individual spacetimes are not necessary to model indeterminism. We share Earman's viewpoint: our characterization of (in)determinism does not require actual branching of spacetimes. Furthermore, and more to the point, \citet[\S3]{earman2008} shows that important theorems assume the Hausdorff property. The theorems listed by Earman exemplify the sense in which the Hausdorff property is considered part of the mathematical machinery necessary for a sensible formulation of GR. Although advocates of branching spacetimes have responded by articulating how appropriate generalizations of the notion of manifolds circumvent such problems to some degree \citep{placekbelnap2012,mueller2013,luc2020interpreting}, it ought to be noted that the non-Hausdorff manifolds are deployed by these authors to fulfil a rather different representational task: the non-Hausdorff structure represents a space of `possible spacetimes' all crowded into a single manifold. 

With the standard representational duties in GR in mind, the Hausdorff property is thus routinely stipulated or implicitly assumed in order to guarantee the smooth workings of the mathematical apparatus. The idea behind taking the status of global hyperbolicity to be that of an \emph{a priori} demand on candidate spacetime models in GR is rather similar: all globally hyperbolic spacetimes are straightforwardly amenable to a well-posed IVF. 

The problem with this approach is that global hyperbolicity is not obviously analogous to the Hausdorff condition, as it turns out that one can formulate a sensible theory in non-globally-hyperbolic spacetimes.\footnote{Cf.\ \citet{Frie04}.} While the latter is widely acknowledged to be a mathematical condition that imposes no physically relevant constraint, global hyperbolicity is thus a substantive assumption. Even if one thinks that determinism is a central physical principle without which no physical theory can operate, enjoining global hyperbolicity thus requires a similarly substantive argument why we ought to accept its \emph{a priori} status. An immediate objection to any resistance to demanding global hyperbolicity would be to question the relevance of non-globally hyperbolic spacetimes, for example spacetimes with CTCs in their maximal extension. Given that we take our universe to have the kinds of features it has, we would not expect that it be best described by a spacetime with CTCs.\footnote{A possible exception to this are Kerr-Newman spacetimes, which contain CTCs in their maximal analytic extension; they describe rotating, charged black holes, which are believed to be present in many regions of spacetime.} The problem with this objection is that if successful, it would only imply that the actual world is deterministic. It would not, however, have any implications as to whether the {\em theory}---GR---is deterministic or not. To the extent to which we are interested in a question about GR, we have to consider the full space of solutions, rather than just some particular solutions that happen to be useful in some particular situations. But this means that we shouldn't simply accept global hyperbolicity as an additional condition which is tacked onto GR in the same way that we ask our manifolds to be Hausdorff spaces. 

Now it seems odd to ask whether we could establish, by observation, that spacetime is Hausdorff. However, one may reasonably hope to justify global hyperbolicity, just {\em because} it is a physically substantive assumption, not by {\em fiat}, but through respectable \emph{a posteriori} observation. The well-known problem is that observations from the confines of a spacetime point, or even a compact region, underdetermine the cosmological model that best describes our actual universe.\footnote{Cf.\ \citet{Glym77} and \citet{Mala77}.} Recently, JB \citet{man09} has proven that, barring one type of pathology, every spacetime is observationally indistinguishable from some other spacetime with different global structure, even fixing all the local properties of the spacetime such as the validity of EFE. The exception includes spacetimes in which there is a point $p \in \mathcal{M}$ such that $I^-(p) = \mathcal{M}$. Not only can all of the spacetime be surveyed from one point in this case, but it can be shown that these spacetimes all contain CTCs. Since spacetimes of this class are thus not globally hyperbolic, this exception is not going to do any work for the defender of \emph{a posteriori} global hyperbolicity: either the spacetimes are already non-globally hyperbolic, or else their global structure is underdetermined.

This result in itself does not completely spoil this line of reasoning.  For even though the global structure of spacetime is generically underdetermined given $I^-(p)$ for any $p$, it could still be the case that \emph{partial} determinations of global structure can be made. In particular, it could be hoped that the global hyperbolicity can be observationally established, at least in principle, while the complete global structure cannot. Alas, \citet{man11} proves that we cannot even make, locally, the determination of whether spacetime is globally hyperbolic.\footnote{This theorem also excludes the same pathological cases.} In his interpretation, this means that although our universe may in fact be globally hyperbolic, we cannot know that it is.\footnote{If our universe {\em fails} to be globally hyperbolic, however, then we may know that \citep{Mala77}.} It shouldn't strike us as particularly surprising that it cannot be determined, from causal information available at a single point, whether or not spacetime is globally hyperbolic, i.e.\ effectively whether the causal signals emanating from everywhere in spacetime must all intersect a global time slice.\footnote{Assuming that causal signals propagate along causal curves, and that these curves exemplify the necessary conditions of inextendibility.} Global hyperbolicity seems to be a property of spacetimes squarely at odds with the project of determining what constrains the global spacetime structure, given our epistemic limitations.

Thus, global hyperbolicity cannot legitimately be established by {\em fiat} and not possibly by observation. Consequently, it assumes the status of a substantive open issue that must be attacked with all the resources the theory can offer. In fact, the {\em (strong) cosmic censorship hypothesis}, roughly the hypothesis that all physically reasonable spacetimes are globally hyperbolic, is widely held to be the most important open issue in the foundations of GR. 
Attempts to prove censorship theorems from the resources of GR, and thus to show that global hyperbolicity can be inferred from weaker assumptions, is the topic of the next section.

\section{Censorship Theorems}
\label{sec:censor}

Hawking motivates censorship theorems as follows \citep[10]{hawpen}:
\begin{quote}
...
[G]lobal hyperbolicity may be a necessity. But my viewpoint is that one shouldn't assume it because that may be ruling out something that gravity is trying to tell us.  Rather, one should deduce that certain regions of spacetime are globally hyperbolic from other physically reasonable assumptions.
\end{quote}
The input assumptions include constraints on the matter sources plugged into EFE and a requirement that the initial data are appropriately `generic'; a proof would show that such initial data yield globally hyperbolic maximal developments.  The name derives from the idea that gravity would act as a `censor', effectively preventing failures of causality from occurring or (in a weaker version of the hypothesis) keeping failures of causality hidden behind the event horizons of black holes. Given a successful proof,  gravity would be telling us that the local existence and uniqueness theorems for GR can be extended globally, ruling out multiple extensions such as the example from \S\ref{sec:determinism}.  In more physical terms, these results would vindicate Penrose's seminal idea that physical effects near a Cauchy horizon generate curvature singularities, preventing extensions.

Although it is difficult to formulate the input assumptions precisely, they are arguably well-motivated by the desire to isolate dynamical effects of GR.  There are two distinct issues. First, one limits consideration to `suitable' matter fields---those that do not themselves develop singularities in Minkowski spacetime.  The idea is to separate singularities due to the matter fields themselves from those due to gravity; it rules out, for example, certain types of fluids because they develop singular shock waves.  The second qualification allows that there may be cases based on highly `special' initial data sets (e.g., with a high degree of symmetry) that lead to a failure of uniqueness.  

This leads to the following formulation of strong cosmic censorship (SCC) \citep{chrusciel1991}.  Given (i) a space of initial data $\mathcal{X}(\Sigma, h_{ab},k_{ab})$, satisfying the Gauss-Codacci constraints, supplemented with appropriate initial data for matter fields; and (ii), a collection of four-dimensional Lorentzian manifolds $\mathcal{S}(M,g_{ab})$ where the metric $g_{ab}$ has some specified differentiability class.  Strong cosmic censorship then states that there is a set $\mathcal{G}(\Sigma,h_{ab},k_{ab})$ in $\mathcal{X}(\Sigma, h_{ab},k_{ab})$ such that the maximal globally hyperbolic development (MGHD) for every data set in $\mathcal{G}$ is inextendible in $\mathcal{S}(M,g_{ab})$. $\mathcal{G}$ is a proper subset of $\mathcal{X}$, as one needs to impose some further conditions in addition to the constraints to ensure that the initial data are themselves `maximal' in an appropriate sense.\footnote{These must at least include a requirment that the initial data are complete or satisfy `fall-off' conditions (such as asymptotic flatness); otherwise initial data specified on an compact region on a spacelike hypersurface in Minkowski spacetime provides a counterexample. See, e.g., \citet{rin09} for further discussion.} We further require that $\mathcal{G}$ must be an open, dense set in an appropriate topology.  This reflects the stability physicists usually require for a well-posed IVF, mentioned at the end of \S \ref{sec:gr}, but faces the same challenges.  Specifying what qualifies as `special' or `generic' initial conditions requires introducing a measure or at least a topology on the space of initial data; but as \citet{wal98} emphasizes, there is not currently a standard measure or topology on the space of initial data, or solutions, that we can appeal to.\footnote{The challenges are similar to those \citet{fle16} identifies with finding a topology on the space of spacetimes.}

There are several purported counterexamples to SCC in the literature, in the form of initial data with MGHD that admit non-isometric extensions given a choice of $\mathcal{S}(M,g_{ab})$.  Yet there are active debates regarding the implications and the appropriate response to these results.

Regarding our example from \S \ref{sec:determinism}, we need not worry about singularities arising from matter fields because Taub-NUT is a vacuum solution. However, stability considerations would presumably rule it out as a counter-example due to the symmetry of the spacetime. There are some results indicating that Cauchy horizons are `special' in the sense of being measure zero in the space of solutions of EFE. For example, \citet{monise} prove that in a particular case (namely, granted that the Cauchy horizon is compact, analytic, and ruled by closed null geodesics) there are symmetries in the neighborhood of the Cauchy horizon.  As a result, this family of solutions would not be expected to fill an open set of initial data in \emph{any} reasonable topology.  \citet{moncrief2020symmetries} have extended their results regarding the existence of symmetries to cover a broader class of spacetimes with Cauchy horizons.  General results along these lines support SCC by showing that the existence of Cauchy horizons does not hold generically, for open sets of initial data, rather than just for specific cases.   

Alongside work showing that purported counterexamples fail the stability requirement, there are a number of theorems establishing global uniqueness for specific choices of $\mathcal{G}$.  In the best case, a proof of SCC would follow for \emph{all} initial data satisfying modest requirements, such as completeness and fall-off conditions.  Current results fall short of that aim, but show that the SCC holds for several quite different families of models.  For example, \citet{chrkla} established the stability of flat spacetime to small perturbations;  roughly, for initial data `close to' Minkowski spacetime, with appropriate decay at spatial infinity, $D(\Sigma)$ is inextendible.  Studies of polarized Gowdy spacetimes (a family of symmetric spacetimes with two independent spacelike Killing fields) have shown that, generically, curvature invariants blow up along curves approaching the Cauchy horizon.  These results establish the existence of a singularity which prevents extensions through the horizon.\footnote{See \citet{isenberg2015strong} for further discussion of Gowdy spacetimes, and \citep{Ande04} for a more general review.}  \citet[][Part IV]{rin09} considers the fate of the SCC within the Bianchi models, and shows that they fall into two classes:  the first exhibit curvature singularities, so that the MGHD cannot be extended, whereas the second class do admit non-isometric extensions of the MGHD.  Yet Ringstr\"om further shows that all of the models in this second case have a further symmetry, suggesting that they would fail to satisfy the stability requirement.   In sum, present results constitute a patchwork, showing that the SCC holds for a variety of specific choices of initial data but falling short of a fully general result.

A different set of issues arises with regards to the choice of $\mathcal{S}(M,g_{ab})$:  in particular, should this set include Lorentzian manifolds where $g_{ab}$ is continuous ($C^0$) but lacks well-defined derivatives?  Several purported counterexamples to SCC assume a positive answer.  \citet{dafermos2017interior}, for example, construct a continuous extension through the Cauchy horizon in Kerr solutions.  Yet the physical significance of merely $C^0$ metrics is not clear.  Second derivatives of the metric appear in the Einstein tensor appearing in the EFE, suggesting that we should impose stronger requirements --- most straightforwardly, that $g_{ab}$ is at least $C^2$, but slightly weaker constraints may allow for EFE to be well-defined in a distributional sense at the Cauchy horizon \citep[][pp. 46-50]{ear95}.  

Furthermore, it should be noted that the choice of $\mathcal{S}(M,g_{ab})$ has implications for the very definition of what it means for a MGHD to be `extendible'.\footnote{We thank a referee for pointing this out to us.} The property of `inextendibility' depends on what one takes the set of relevant models to be: one and the same Lorentzian manifold may be inextendible with respect to one background collection of models, but extendible with respect to another. Restrictions to `physically reasonable' models, $\mathcal{P} \subset \mathcal{S}$, impact assessments of modality. \citet{manchak2018} argues that for every non-trivial choice of $\mathcal{P}$, it is not the case that every $\mathcal{P}$-inextendible $\mathcal{P}$-model is inextendible in $\mathcal{S}$. This obviously has a direct bearing on debates regarding the SCC, given that alleged counterexamples may fall outside some chosen collections $\mathcal{P}$.   

Decades of effort have led to much sharper formulations of the SCC, yet the proof of a general version of the SCC has remained elusive.  But perhaps the results that have already been achieved are sufficient; in what sense would a patchwork quilt of stability results answer our original question regarding whether GR is deterministic?  In the best case scenario, the sets $\mathcal{G}$ for which SCC holds would cover all the solutions used in black hole physics or cosmological modeling.  This would involve a much more parochial sense of physical possibility, as it would apply directly to only a subset of the full solution space of GR.  Although the question of whether classical GR is deterministic, full stop, would remain unresolved, such results would provide assurance that no failure of determinism would crop up in the applications of GR to physical systems covered by results making up the quilt.

\section{Incompleteness of GR}
\label{sec:beyondgr}

Nobody expects GR to be the last word on gravitation. Accordingly, an advocate of global hyperbolicity may draw hope from the fact that non-globally hyperbolic spacetimes may be incompatible with the quantum. As it turns out, however, the direction the physics literature indicates in this respect is far from unambiguous. For instance, considerations in semi-classical quantum gravity (QG) pull both ways. On the one hand, the usual point-wise energy conditions all fail for quantum fields. Thus, quantum effects may undermine efforts to establish cosmic censorship from assumptions concerning what energy conditions the energy-matter content of the universe may reasonably be expected to satisfy. On the other hand, quantum effects may effectively act as barriers at Cauchy horizons, cutting off spacetime extensions beyond the globally hyperbolic domain.\footnote{For a survey of related results, cf.\ \citet[\S5]{earsmewut} and \citet[\S7]{smewut}.}

While results in semi-classical QG pull both ways, an attempt to formulate a QFT on curved spacetimes may  provide respectable grounds to stipulate global hyperbolicity from the get-go. A standard recipe for obtaining a natural algebra $\mathfrak{A} (\mathcal{M}, g_{ab})$ of observables for a quantum field defined over a spacetime $\langle \mathcal{M}, g_{ab}\rangle$ only succeeds if the spacetime is globally hyperbolic. But the defender of global hyperbolicity will not find much solace in this, for two reasons. First, since QFT on curved background spacetimes does not consider the backreaction of the quantum fields on the spacetime it has no prayer of being a fundamental theory; at best, it is a 0th approximation to a full quantum theory of gravity that can be useful to model the weak-field regime. Second, it seems possible that the conditions necessary for the construction of the algebra of observables may be weakened in such ways as to enable the procedure to apply to non-globally hyperbolic spacetimes. To be sure, some of these relaxations are problematic; but it is far from clear how sacrosanct the condition of global hyperbolicity really is.\footnote{For further details, cf.\ Earman et al.\ (2009, \S 6).}

Of course, only a full theory of QG will be able to give more authoritative answers. But since there exists such a plethora of competing approaches, it is currently impossible to draw many general conclusions from research in QG. One possibility that must be taken into serious consideration, however, is the suggestion offered, in one form or another, by many approaches to QG that spacetime fails to be fundamental, but instead emerges from an underlying, non-spatio-temporal stucture \citep{hugwut}. In this case, it may become impossible to straightforwardly connect global hyperbolicity, in the sense defined above, to this fundamental structure. Assuming, however, that this can be done---as it has to in order for the theory to have a meaningful classical limit---, we see in principle three different manners in which the issue of global hyperbolicity may play out in QG. First, it may be postulated by {\em fiat} (as discussed in \S\ref{sec:gh}) as a transcendental condition for the approach. This is unproblematic as long as it is understood that the assumption ultimately requires either independent support or else justification {\em ex post facto} from the success of the theory based on it. Second, global hyperbolicity, or what amounts to it in the appropriate classical limit, may be derived from the resources of the approach itself, or from conditions solidly grounded in considerations concerning what is physically reasonable drawing, perhaps, on the resources of other theories. As a third possibility, it may turn out that global hyperbolicity or its quantum analogue is not vindicated by the fundamental theory. In this latter case, it is hard to imagine a sound reason for insisting on global hyperbolicity at the classical level. 

The third possibility could be borne out in string theory. There are strong indications that at least in five-dimensional supersymmetric gravity, a close relative to string theory, analogues of spacetimes with CTCs exist and thus violations of global hyperbolicity may occur \citep{gau03}. Since solutions of the five-dimensional theory can easily be worked up to solutions for ten- and eleven-dimensional supergravity, the situation in string theory is arguably not much different from the one in GR. The second option above would amount to an attempt to prove an analogue of cosmic censorship in full QG. Finally, several approaches, among them causal sets and loop quantum gravity (LQG), go for the first alternative, i.e., they essentially impose global hyperbolicity or a close cousin {\em ab initio}.\footnote{Although it is not obvious what the discrete analogue of global hyperbolicity would be, it seems natural to think that an adequately transposed condition holds in the causal sets approach. Its basic assumption of acyclicity certainly rules out CTCs.} The canonical quantization procedure applied in LQG simply requires global hyperbolicity in order to get traction. However, an alternative viewpoint conceives of LQG not as primarily a theory concerned with the global structure of spacetime, but rather as offering a quantum theory of small patches of spacetime as may be measured in a laboratory. This alternative would view LQG not as cosmological, but instead as concerned with local physics \citep{wut20}. 
We believe that it is unfortunate that this viewpoint is rarely taken seriously, but submit that according to it, LQG could not be a fundamental quantum theory of spacetime capable of accounting for the global structure of spacetime. 

As is evident, once again, the field is simply too rich to draw robust general conclusions concerning the fate of determinism. But even though this may come as a disappointment to some, we believe that the many foundational lessons that can be gleaned from the field make the endeavour more than worthwhile.

\section{Conclusions}
\label{sec:conc}

In our investigation of the fate of determinism in GR and beyond, we have explored three different approaches to justifying the imposition of global hyperbolicity. The first approach aimed to justify this condition directly. Global hyperbolicity is a \emph{global} property of spacetime, and it generally cannot be established via observations of a local region.  So it appears that the condition has to be treated as an \emph{a priori} constraint, and we considered, but rejected, the possibility of treating it as a mathematical constraint transcendentally necessary for the formulation of a sensible theory. The second approach seeks to establish `censorship theorems', showing that global hyperbolicity holds for some subset of spacetimes that satisfy weaker, physically motivated requirements. While there are some promising results, this approach is a work in progress and it remains genuinely open whether it will succeed in establishing global hypobolicity, at least for some relevant classes of models. Finally, a third approach insists on the need to consider the limits of applicability of classical GR and its inter-connections with other theories, such as QFT on curved spacetimes, semi-classical and full QG. Although more speculative, this approach bears the promise to elucidate the relationship between GR and more fundamental theories towards quantum gravity and thus may have significant heuristic value for articulating a theory of quantum gravity. 

Our discussion of the fate of determinism in GR required a decision regarding the concept of determinism itself. The standard Laplacian notion according to which the state of the world at a time, together with the laws of nature, jointly determine the state of the world at all times naturally translates into the requirement that the (geometric and matter) data on a global time slice $\Sigma$ uniquely extend to all of spacetime. The existence and uniqueness of such an extension thus required for the satisfaction of determinism as usually envisioned by philosophers does not, however, suffice for a well-posed IVF as sought after by physicists. In addition, solutions must continuously depend on the initial data. Although technically involved, this continuity condition, if added, rules out `unphysical' failures of determinism due to exquisitely fine-tuned initial data. Thus, the question of determinism in GR raises not only deep and fruitful philosophical and foundational issues, but it also serves to bring out a contrast between philosophical interests and the demands of applied mathematics. 

We hope to have highlighted, once again, the value of assessing the fate of determinism in modern physics.  There is simply no quick answer to be had, as it depends on a plethora of subtle conceptual and technical issues in the foundations of the theories considered. But this is what makes determinism such a central topic in the exploration of physical theories in general, and of GR in particular.

\bibliographystyle{plainnat}
\bibliography{determinism}

\end{document}